\documentstyle[aps]{revtex}
\begin{document}
\title{An Inherently Quantum Computation Paradigm: NP-complete=P Under the
Hypothetical Notion of Continuous Uncomplete von Neumann Measurement}
\author{Giuseppe Castagnoli}
\address{Information Communication Technology Dept., \\
Elsag Bailey, Via Puccini 2, 16154 Genova, Italy}
\date{\today }
\maketitle

\begin{abstract}
The topical quantum computation paradigm is a transposition of the Turing
machine into the quantum framework. Implementations based on this paradigm
have limitations as to the number of: qubits, computation steps, efficient
quantum algorithms (found so far). A new exclusively quantum paradigm (with
no classical counterpart) is propounded, based on the speculative notion of 
{\em continuous} uncomplete von Neumann measurement. Under such a notion,
NP-complete is equal to P. This can provide a mathematical framework for the
search of implementable paradigms, possibly exploiting particle statistics.
\end{abstract}

\section{Introduction}

During the Helsinki meeting on quantum computation and communication
(September 26-28, 1998), some participants explicitly asserted that full
fledged quantum computation is completely out of the current course of
research (in my understanding: N. Gisin, S. Harroche, R. Landauer, A.
Zeilinger). Going beyond 10 qubits (even in the most optimistic estimates)
and a very limited number of computation steps would be out of reach for
today's technology and paradigm. Therefore, new ideas and, in particular,
new computation paradigms would be needed (A. Ekert, the author).

The following is an extended version of part of my speech. It concerns an
alternative quantum computation paradigm, completely speculative for the
time being.

This paradigm appears in former works of the author and others$^{\left[ 1-6%
\right] }$, where it is somewhat buried inside implementation attempts. Here
I will explain its bare principle, by resorting to the speculative notion of 
{\em continuous} uncomplete von Neumann measurement.

\section{A critique of Turing-quantum computation}

Until now, the topical quantum computer is a transposition of the classical
Turing machine into the quantum framework. On one side of the coin, there
are the well-known revolutionary results, on the other side, severe
limitations are being encountered. One could think that the very importance
of the results obtained might trap research inside a limited horizon.

In the first place, we should clearly acknowledge the boundaries implicit in
the notion of Turing machine computation.

Before going to this, it is useful to introduce a special perspective. All
efficient quantum algorithms found so far (more efficient than their
classical counterparts as far as we know) do nothing but solving systems of
simultaneous Boolean equations. In fact all NP and NP-complete problems%
\footnote{%
The search problem, belonging to P, will be discussed at the end.} can be
solved by a Turing machine in a number of elementary computation steps which
has an upper bound. Because of this feature, they can be represented by a
system of simultaneous Boolean equations, whose size is polynomial in
problem size in the case of NP and NP-complete problems. It is convenient to
have in mind one system of this kind and its network representation (fig. 1):

\begin{tabular}{l}
$x_{3}=Nand(x_{1},x_{2})$ \\ 
$x_{6}=Nand(x_{4},x_{5})$ \\ 
$x_{1}=x_{5}$ \\ 
$x_{2}=x_{6}$ \\ 
$x_{3}=x_{4}$%
\end{tabular}

\qquad \qquad \qquad \qquad \qquad \qquad \qquad \qquad \qquad \qquad Fig. 1

This system implicitly {\em defines} the solutions: $x_{1}=x_{5}=0$, $%
x_{2}=x_{6}=1$, $x_{3}=x_{4}=1$, and two others. Of course, it does not
describe a physical system that constructs the solutions.\footnote{%
If fig. 1 network were implemented as an electronic circuit, it would start
oscillating like a flip-flop, because of the time delays and the lack of
simultaneity of the classical world. In general, the probability of its
settling down in a solution decreases exponentially with network size.} In
other words, the system of simultaneous equations (i.e. the problem) is {\em %
the definition of an object} (the solutions) but it does not map directly
the way of constructing the object. In order to construct it, one should
scan the possible Boolean variable assignments and, for each assignment,
check whether it satisfies all gates and wires, thus following a very
indirect way. This is a {\em sequential} task that a Turing machine can
perform by undergoing a suitable time evolution. Summing up, while the
definition of the object is a-sequential and a-temporal, its classical
construction needs to be sequential and temporal.

On the contrary, the computation paradigm I am going to expound will make a
limited use of sequentiality. It will rely on the notion that, in the
quantum framework, there could be {\em identity} between defining and
constructing. By the way, this resembles the description-action
identification asserted by D. Finkelstein: there are only actions$^{\left[ 7%
\right] }$.

Going back to the boundaries of Turing machine quantum computation, one can
see that:

\begin{enumerate}
\item  a computational efficiency unknown in the classical framework is
achieved by exploiting {\em some} exclusively quantum features $-$ those
involved in multiparticle interference. See for example ref. [8]. Whereas
the basic computation paradigm remains a classical one.

It is characterized by the {\em logical and physical sequentiality of
computation}. For example, let us consider the task of computing a function
given the argument. The argument is the input and the result of computation
is the output, necessarily separated from the input by a non-zero time
interval.

An alternative characterization of sequentiality lies in the fact that {\em %
a reversible Boolean network appears in the time-diagram} {\em of\ the
computation process}.

I will show that such a classical notion of sequential computation, inherent
in the Turing paradigm, must be in some measure {\em given} {\em up}, if one
wants to achieve any computation speed up in the quantum framework.

\item[2.]  Until now, the quantum Turing machine exploits an {\em uncomplete}
set of exclusively quantum features: there is no reason whatsoever to think
that a quantum computer based on such features can efficiently simulate
phenomena involving {\em other} exclusively quantum features, like, for
example, particle (boson, fermion, anyon, ...) statistics. Interestingly, in
his 1982 paper$^{\left[ 9\right] }$, Feyman wrote: ``I'm not sure whether
Fermi particles could be described by such a system [in my understanding,
today's ``universal'' quantum computer]''. Parenthetically, the problem of
the simulation of fermion statistics with today's ``universal'' quantum
computer is addressed in ref. [10-11, among others].

I will conjecture that a more radical detaching from sequentiality implies
the exploitment of new exclusively quantum features, possibly particle
statistics.
\end{enumerate}

From a methodological standpoint, it is legitimate to think that the above 
{\em inherent} limitations of Turing quantum computation might have
something to do with the severe limitations encountered by the
implementations: limited number of qubits, computation steps, and efficient
algorithms. Moreover, the Turing-quantum paradigm is suspected of being
inherently unable to solve an NP-complete problem in polynomial time$^{\left[
12\right] }$. Perhaps we are trying to force the quantum nature into a
scheme un-congenial to it.

This work explores the possibility of exploiting the quantum nature in an
alternative and perhaps more congenial way. Although the computation
paradigm propounded will be completely speculative, it might help in
``thinking out of the box'' of the Turing paradigm.

\section{The relaxation of classical sequentiality in the quantum-Turing
paradigm}

Classical sequentiality implies that there must be a time interval between
the input and the output and that the output is a function of the input
alone. I will show that this is no more the case in quantum Turing
computation.

Let us consider Simon's algorithm$^{\left[ 13-14\right] }$. At some stage of
it, at time say $t_{f}$, the computer state is:

\begin{equation}
\left| \psi \left( t_{f}\right) \right\rangle =\sum_{x}\left| x\right\rangle
_{a}\left| f\left( x\right) \right\rangle _{b},
\end{equation}

\noindent where $a$ ($b$) is the register containing the argument (the
function). This function is periodic and defined over two periods. State (1)
is naturally a function of the input (the preparation), which it follows in
time.

At the next stage of the algorithm, it is convenient to think that we
measure the content of register $b$ in state (1), obtaining $f\left(
x\right) =k$, some constant value. There are two values of $x$ such that $%
f\left( x\right) =k:$ $x=\overline{x}$ and $x=\overline{x}+p,$ where $p$ is
the period. Consequently state (1) changes into 
\begin{equation}
\left| \psi \left( t_{f+}\right) \right\rangle =\left( \left| \overline{x}%
\right\rangle _{a}+\left| \overline{x}+p\right\rangle _{a}\right) \left|
k\right\rangle _{b}
\end{equation}

\noindent Further ingenuity and operations allow to extract $p$ out of the
superposition (2). But the quantum trick, giving the ``speed up'', has
already been done$^{\left[ 6\right] }$.

Now, the essential point is that state (2) {\em is a function of both the
input and the output}, in this case the outcome of measurement.

The ``wave function collapse'', from (1) to (2), is not necessarily located
after $t_{f}$. According to von Neumann and Wigner (among others) it can be
retrodicted to any time $t$ between preparation and measurement, provided
that the deterministic evolution of the collapsed state at $t$, gives
eventually the state after measurement at $t_{f+}$. Therefore, we can place
the collapse before $t_{f}$. Thus $\left| \psi \left( t_{f}\right)
\right\rangle $ is either (2) or (1) whether or not the collapse has been
taken into account. In the former case, we have a deterministic evolution
from the result of measurement to $t_{f}$; this evolution undergoes back in
time the unitary transformations of the conventional evolution. In the
latter, we have the conventional evolution from preparation to $t_{f}$. This
means that measurement changes a quantum state which was a function of the
input into a state which is a function of both the input and the output.
This {\em quantum} {\em violation of sequentiality} justifies the ``quantum
speed up'', as shown in ref. [6].

The exclusively quantum computation paradigm I am going to expound is in a
way an extrapolation of the quantum mechanism discussed above. We will be
dealing with a quantum computation state that changes in function of
conditions placed both in its immediate past and future, under a {\em %
speculative}, {\em continuous} uncomplete von Neumann measurement.

\section{An exclusively quantum computation paradigm}

It is convenient to use another way of representing a general system of
simultaneous Boolean equations: the feedback loops of fig. 1 are substituted
by the condition that the output of an open Boolean network has a
pre-assigned value, conventionally 1 (fig. 2).

\begin{tabular}{l}
$y=f\left( x_{1},x_{2},...,k_{1},k_{2},...,\right) $ \\ 
$k_{1}=0$ \\ 
$k_{2}=1$ \\ 
$...$ \\ 
$y=1$%
\end{tabular}
\qquad \qquad \qquad \qquad

\qquad \qquad \qquad \qquad \qquad \qquad \qquad \qquad \qquad \qquad Fig. 2

In fig. 2, part of the input, $k_{1}$, $k_{2},...$, as the output $y$, are
pre-established, the other part of the input, $x_{1},x_{2},...,$ is
``unknown''. $f$ is a general Boolean function. The problem is whether there
is an assignment to the unknown part of the input, such that the output is $%
y=1$. This is the well-known SAT problem, which is NP-complete. Input and
output have a purely logical meaning; they mean argument and function. As a
matter of fact, they will be {\em coexisting} qubits.

I shall give a preliminary outline of the model first. This will not be
accurate, but gives the line of thinking. Let 
\begin{equation}
{\cal H}=span\left\{ \left| 0\right\rangle _{x_{1}}\left| 0\right\rangle
_{x_{2}}...\left| 0\right\rangle _{k_{1}}\left| 0\right\rangle
_{k_{2}}...\left| 0\right\rangle _{y},...,\left| 1\right\rangle
_{x_{1}}\left| 1\right\rangle _{x_{2}}...\left| 1\right\rangle
_{k_{1}}\left| 1\right\rangle _{k_{2}}...\left| 1\right\rangle _{y}\right\} ,
\end{equation}

\noindent be the Hilbert space the independent qubits,

\begin{equation}
{\cal H}^{c}=span\left\{ 
\begin{array}{c}
\left| 0\right\rangle _{x_{1}}\left| 0\right\rangle _{x_{2}}...\left|
0\right\rangle _{k_{1}}\left| 1\right\rangle _{k_{2}}...\left| f\left(
0,0,...,k_{1}=0,k_{2}=1,...\right) \right\rangle _{y},..., \\ 
\left| 1\right\rangle _{x_{1}}\left| 1\right\rangle _{x_{2}}...\left|
0\right\rangle _{k_{1}}\left| 1\right\rangle _{k_{2}}...\left| f\left(
1,1,...,k_{1}=0,k_{2}=1,...\right) \right\rangle _{y}
\end{array}
\right\} ,
\end{equation}

\noindent be a constrained subspace of ${\cal H}$, whose basis vectors
satisfy the gate logical constraint and the pre-assigned input, and $P_{f}$ $%
\left( P_{f}^{2}=P_{f}\right) $ be the projector from ${\cal H}$ on ${\cal H}%
^{c}$.

We suppress the constraint $y=1$ and give an arbitrary assignment to the
unknown part of the input, obtaining (in polynomial time), say, $y=0$.
Obtaining $y=1$ would mean solving the problem by sheer luck, which is
disregarded.

We prepare the independent qubits in such an assignment. Then we rotate only
qubit $y$ from $\left| 0\right\rangle _{y}$ to $\left| 1\right\rangle _{y}$,
under continuous uncomplete von Neumann $P$ measurement of the entire set of
qubits. We will see that, correspondingly, the input changes to a value such
that the output is $y=1$ (provided there is a solution). This is in a way an
extrapolation of Simon's algorithm ``trick'' [changing from state (1) to
state (2), because of measurement of the $b$ register only].

In order to show this, it is useful to clarify first the notion of
continuous measurement. This is better done by considering a simple
(reversible) identity gate\footnote{%
When gate input and output qubits coexist, one can use indifferently a
logically reversible or irreversible gate. We should note that in sequential
computation there can be states which map a logically irreversible gate
between the coexisting qubits of the register, which, at some stage,
contains both the output of computation and the memory of the input.}.

\begin{center}
Fig. 3
\end{center}

The two qubits $x$ and $y$ coexist and are independent. It should be noted
that measurement under projector $P$ concerns a joint property of the two
qubits, whether or not their overall state satisfies the identity gate. $P$
measurement yields an outcome belonging to either

\begin{equation}
{\cal H}^{c}=span\left\{ \left| 0\right\rangle _{x}\left| 0\right\rangle
_{y},\left| 1\right\rangle _{x}\left| 1\right\rangle _{y}\right\} ,
\end{equation}
\[
\text{or to }\overline{{\cal H}}^{c}=span\left\{ \left| 0\right\rangle
_{x}\left| 1\right\rangle _{y},\left| 1\right\rangle _{x}\left|
0\right\rangle _{y}\right\} . 
\]

\noindent The former subspace satisfies the gate, the latter does not.

We ask the question whether there is a way of applying $P$ measurement to
obtain that qubit $x$ follows a rotation applied only to qubit $y$. The
answer is NO if measurement is {\em intermittent}, even in the limit of
infinite frequency (Zeno effect). The answer is YES if measurement is {\em %
continuous} to start with.

We shall consider the former case first. We should start from a state
satisfying the gate, say $\left| 0\right\rangle _{x}\left| 0\right\rangle
_{y}$ for simplicity. Time is split into small consecutive intervals $\Delta
t$ (fig. 3). \ During each interval $k$ an $\omega t$ rotation is applied,
with $k\Delta t<t<(k+1)\Delta t$. At the end of the first interval we have
the state:

\[
\left| \psi \left( 0\right) \right\rangle =\left| 0\right\rangle _{x}\left|
0\right\rangle _{y}\rightarrow \left| \psi \left( \Delta t\right)
\right\rangle =\cos \omega \Delta t\left| 0\right\rangle _{x}\left|
0\right\rangle _{y}+\sin \omega \Delta t\left| 0\right\rangle _{x}\left|
1\right\rangle _{y}. 
\]

\noindent Then the two qubits are measured under $P$. If $\Delta t$ is
infinitesimal, with certainty the outcome of measurement is $\left|
0\right\rangle _{x}\left| 0\right\rangle _{y}$ back again. This holds for
any time $t$ (for any number of repetitions of the rotation-measurement
process) in the limit $\Delta t$ $\rightarrow 0$. This is of course the Zeno
effect, which freezes the evolution in its initial state. It depends on the
fact that, in the quantum framework, there are both probability amplitudes
and probabilities, which are the squared modulus of the former ones. The
probability that the outcome of measurement violates the gate is higher
order infinitesimal ($\sin ^{2}\omega \Delta t$) and can be disregarded.

Here, it is important to note that during the entire time interval $\Delta t$
the gate is violated because of the presence of $\sin \omega t\left|
0\right\rangle _{x}\left| 1\right\rangle _{y}$, with $0<t<\Delta t$. The
percentage measure of time during which it is violated is 100\%: it is not
violated in two points in time, it is violated during the entire time
interval comprised between the two points. Naturally such a percentage
measure remains unaltered in the limit $\Delta t$ $\rightarrow 0$. To sum
up, the time measure during which the gate is violated coincides with any
elapsed time $t$.

Let us now consider {\em continuous} measurement. If measurement is
continuous to start with, {\em not} in the limit $\Delta t$ $\rightarrow 0,$
the gate can never be violated. If we start from an initial state that
satisfies the gate, continuous measurement must always keep the state inside 
${\cal H}^{c}$. The evolution under continuous measurement must satisfy the
simultaneous conditions$^{\left[ 3,5\right] }$:

\begin{enumerate}
\item[i)]  $\left| \psi \left( 0\right) \right\rangle =\left| 0\right\rangle
_{x}\left| 0\right\rangle _{y},$

$\forall t>0:$

\item[ii)]  $P\left| \psi \left( t\right) \right\rangle =\left| \psi \left(
t\right) \right\rangle ,$ where $P$ is the projector from ${\cal H}$, the
Hilbert space of the two independent qubits, on ${\cal H}^{c}$ given by eq.
(5),

\item[iii)]  the distance between the vectors before and after measurement
is minimum,

\item[iv)]  $Tr_{x}\left| \psi \left( t\right) \right\rangle \left\langle
\psi \left( t\right) \right| =\cos ^{2}\omega t\left| 0\right\rangle
_{y}\left\langle 0\right| _{y}+\sin ^{2}\omega t\left| 1\right\rangle
_{y}\left\langle 1\right| _{y},$ where $Tr_{x}$ means partial trace over $x.$
\end{enumerate}

\noindent At any time $t$, $\left| \psi \left( t\right) \right\rangle $ is a
generic normalized vector of ${\cal H}$; if $t_{1}\neq t_{2}$, $\left| \psi
\left( t_{1}\right) \right\rangle $ and $\left| \psi \left( t_{2}\right)
\right\rangle $ are two {\em independent} generic vectors of ${\cal H}$. See
also ref. [5].

It should be noted that the above conditions yield a very direct mapping of
the gate, under any possible transformation of its state. We can see that
sort of identity between definition and construction, discussed in Section
I. However, we are of course in a speculative situation.

It can be seen that conditions (i), (ii), and (iv) yield

\begin{equation}
\left| \psi \left( t\right) \right\rangle =\cos \omega t\left|
0\right\rangle _{x}\left| 0\right\rangle _{y}+e^{i\delta }\sin \omega
t\left| 1\right\rangle _{x}\left| 1\right\rangle _{y}.
\end{equation}

Condition (iii) keeps the phase $\delta $, arbitrarily chosen once for all
if the initial state is $\left| 0\right\rangle _{x}\left| 0\right\rangle _{y}
$, frozen throughout the evolution,\ thus for $t>0$.

It can be seen from equation (6) that qubit $x$ identically follows the
rotation of qubit $y$, under this continuous uncomplete measurement.

Interestingly, evolution (6)\ is {\em driven }by both{\em \ the initial
condition }(i){\em \ and the final condition} that the result of $P$
measurement is a vector belonging to ${\cal H}^{c}$ and satisfying
conditions (iii)\ and (iv). Conditions (i) through (iv) make up a peculiar
variational problem. In a sense, it is Simon's ``trick'' brought to the
continuum.

If instead we started from the initial state 
\begin{equation}
\left| \psi \left( t\right) \right\rangle =\cos \varphi \left|
0\right\rangle _{x}\left| 0\right\rangle _{y}+e^{i\delta }\sin \varphi
\left| 1\right\rangle _{x}\left| 1\right\rangle _{y},
\end{equation}

\noindent with $\delta $ pre-established as an initial condition, the
evolution would have been

\begin{equation}
\left| \psi \left( t\right) \right\rangle =\cos \left( \varphi +\omega
t\right) \left| 0\right\rangle _{x}\left| 0\right\rangle _{y}+e^{i\delta
}\sin \left( \varphi +\omega t\right) \left| 1\right\rangle _{x}\left|
1\right\rangle _{y}.
\end{equation}

\noindent Evolution (8) avoids the difficulty of evolution (7), namely the
appearance ``out of the blue'', of the term $\left| 1\right\rangle
_{x}\left| 1\right\rangle _{y}$, with an infinitesimal amplitude and a
random phase. This resembles a state vector reduction reversed in time, and
may not be palatable. Evolution (8) will serve our needs as well.

Now we go back to the general gate of fig. 2. Each unknown input $i$ is
prepared in the superposition $\frac{1}{\sqrt{2}}\left( \left|
0\right\rangle _{i}+\left| 1\right\rangle _{i}\right) $, unlike the
``inaccurate'' outline. This can be propagated in polynomial time to the
(unconstrained) output by using a ``conventional'' quantum Boolean network,
such that it keeps, in its output, the memory of the input. Such coexisting
output qubits are of course the logical input and output appearing in fig.
2. We remain with an independent set of qubits prepared in the state:

\begin{equation}
\left| \Psi \left( 0\right) \right\rangle =\cos \vartheta \sum_{i}\alpha
_{i}\left| u_{i}\right\rangle _{x}\left| 0\right\rangle _{y}+\sin \vartheta
\sum_{j}\alpha _{j}\left| u_{j}\right\rangle _{x}\left| 1\right\rangle _{y}.
\end{equation}
with $\sum_{i}\left| \alpha _{i}\right| ^{2}=\sum_{j}\left| \alpha
_{j}\right| ^{2}=1;$ $\left| u_{i}\right\rangle _{x}$, $\left|
u_{j}\right\rangle _{x}$ denote tensor products of all qubit eigenstates but
qubit $y$. $i$ ($j$) ranges over the set of input arguments whose function
is $y=0$ ($y=1$). If the problem is hard to solve, $\sin \vartheta $ is
exponentially smaller (in qubits number) than $\cos \vartheta .$

Now we apply the same manipulations already applied to the identity gate:
the initial state becomes the right hand of eq. (9), $P$ becomes the
projector from ${\cal H}$ given in (3) on ${\cal H}^{c}$ given in (4),
pertaining the general Boolean gate. \footnote{%
By the way, any Boolean gate can be made of elementary Nand gates and wires
(labeled by $h)$. Interestingly, $P$ is the product of the projectors from $%
{\cal H}$ on ${\cal H}_{h}^{c}$, the Hilbert space spanned by the basis
vectors satisfying gate or wire $h:$ $P=\Pi _{h}P_{h}.$ All such projectors
are diagonal in the qubit basis, thus pairwise commuting. We could speak as
well of continuous uncomplete measurement under the commuting projectors $%
P_{h}$.}

With an $\omega t$ rotation of qubit $y$ ranging from 0 to $\pi /2$, state
(9), submitted to (updated) conditions (i) trough (iv), changes into

\begin{equation}
\left| \Psi \left( \frac{\pi }{2}\right) \right\rangle =\sin \vartheta
\sum_{i}\alpha _{i}\left| u_{i}\right\rangle _{x}\left| 0\right\rangle
_{y}+\cos \vartheta \sum_{j}\alpha _{j}\left| u_{j}\right\rangle _{x}\left|
1\right\rangle _{y},
\end{equation}

\noindent as it can be readily checked. Now the probability of measuring $%
\left| 0\right\rangle _{y}$ is exponentially small. A solution is found
almost certainly, provided there is one. However, if the result of
measurement is not a solution, this can be checked in polynomial time. By
repeating the overall process for a sufficient number of times, one can
ascertain with any desired confidence level whether the network admits a
solution.

Therefore, it is NP-complete=P under this {\em continuous} uncomplete von
Neumann measurement. Whether continuous measurement is physical and,
possibly, what are the resources required to implement it, remain open
problems.

\section{Discussion}

The notion of continuous measurement is physically plausible, in the usual
sense that it is in agreement with some physical laws and does not seem to
contradict any other laws. Therefore it can be argued that we have given a
plausible physical scheme under which NP-complete=P. The time required to
find a solution (if any) is the time required to rotate one independent
qubit by $\pi /2$. Conventionally, this is just one computation step. Such a
scheme might help in thinking out of the box of the conventional
Turing-quantum paradigm.

In former work$^{\left[ 3-5\right] }$, it has been conjectured that
continuous measurement appears in nature in the form of particle (fermion,
boson, anyon, ...) statistics. Particle statistics can as well be seen as
continuous measurement under the projector on the symmetric Hilbert subspace.

From being a passive constant of motion that does nothing to a unitary
evolution, particle statistics should assume an active role, it should
``shape'' the evolution of which it becomes a constant of motion. One should
look for coupling qubits by means of symmetries purely induced by particle
statistics. The exotic kinds like anyon statistics, would provide a broader
range of investigation. For example, we plan to investigate the possibility
that the NOT\ and AND gate given in ref. [2], where qubit coupling is
obtained by (respectively) fermion and anyon statistics, behave as foreseen
in Section IV, under rotation of one qubit.

It has been observed that finding a solution by amplifying its amplitude, as
we did by transforming state (9) into state (10), resembles Grover search
algorithm$^{\left[ 14,15\right] }$, which has been demonstrated to be optimal%
$^{\left[ 16\right] }$, and gives no exponential speed up. However, our
(hypothetical) exponential speed up applies to an NP-complete problem,
whereas the search problem is P (polynomial). The system of Boolean
equations specifying it\footnote{$%
x_{0}=0,x_{1}=0,...,x_{k}=1,...,x_{N-1}=0,x_{N}=0.$} has the same dimension
of the classical search process and is already solved. Therefore, the
current scheme gives no advantage in a completely explicit search problem.

Thanks are due to A. Ekert, D. Finkelstein, M. Rasetti, and V. Vedral for
useful discussions.

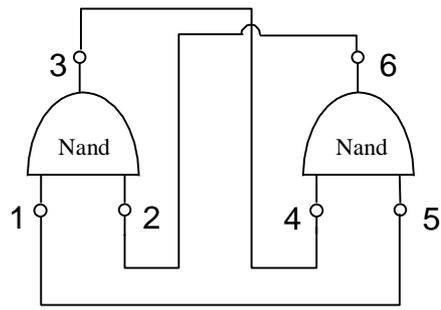
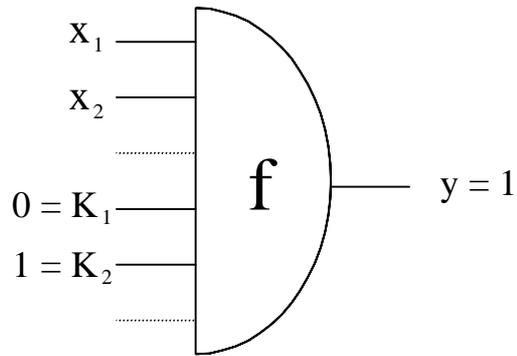
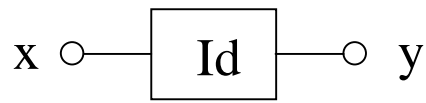
\end{document}